\documentclass[twocolumn,aps,prd]{revtex4}
\newcommand{\be}{\begin{equation}}
\newcommand{\ee}{\end{equation}}
\newcommand{\bea}{\begin{eqnarray}}
\newcommand{\eea}{\end{eqnarray}}

\newcommand{\ta}{\tilde\alpha}
\newcommand{\tb}{\tilde\beta}

\newcommand{\bm}{\bar\mu}
\newcommand{\bn}{\bar\nu}

\newcommand{\la}{\lambda}

\newcommand{\e}{\eta}

\newcommand{\om}{\omega}

\newcommand{\sss}{\sigma}
\newcommand{\ssb}{{\overline{ \sigma}}}
\newcommand{\sq}{\sqrt{2}}
\newcommand{\sqs}{\sqrt{6}}
\newcommand{\sqt}{\sqrt{3}}
\newcommand{\sqf}{\sqrt{5}}
\newcommand{\sqtt}{\sqrt{3\over 2}}
\newcommand{\os}{\overline\Sigma}
\newcommand{\s}{\Sigma}
\newcommand{\Sigb}{{\overline\Sigma}}
\newcommand{\oot}{\overline {126}}
 \newcommand{\bh}{\bar h}
\newcommand{\bt}{\bar t}

\newcommand{\ovl}{\overline}
\begin{document}
\title{On the Consistency of MSGUT Spectra }

\author{ Charanjit S. Aulakh }

\affiliation{ Dept. of Physics, Panjab University,
 Chandigarh, India 160014}

\begin{abstract}
We show explicitly that, once convention dependent phases
are properly accounted for,  the
 mass  spectra  evaluated by us
in \cite{msgt04} satisfy the
Trace,   SU(5) reassembly and Goldstone counting
consistency checks. Furthermore proper phase accounting
shows that the transposition symmetry called ``Hermiticity"
  will be manifest only up
to signs arising from the product of six phase
factors which must be reinserted if the symmetry
is to be verified also as regards signs.
Thus the claims of  hep-ph/0405300 and hep-ph/0412348
 concerning the inconsistency  of our results are completely mistaken.
 The Chiral multiplet spectra of the two calculations are
 equivalent.  However our method also  gives all gauge and
  gauge chiral  spectra as well as a MSSM
 decomposition of all SO(10) MSGUT couplings,
 for both tensors and spinors,   which are
 unavailable, even in principle,
 using the methods of the above papers.

\end{abstract}
\maketitle
\section{Introduction}
The supersymmetric $SO(10)$ GUT
based on the $210-126-\overline{126}$ Higgs system
which was proposed in
refs.\cite{aulmoh,ckn} is of late basking in the attentions
of the GUT model builders' community. There are
 two principal reasons
for this recent revival- apart from the general favour
 enjoyed by SO(10) GUTs  since the
discovery of neutrino oscillations. The first was the
the realization \cite{abmsv03}
that this SO(10) Susy GUT is in fact the {\it{ minimal}}
(Susy)   GUT (MSGUT) compatible with all known data.
Secondly, although the proposal of
\cite{bm93} to {\it{predictively}}
fit all charged fermion and neutrino masses using only
  the $10-\oot$ Fermion Mass (FM)
Higgs system  initially seemed\cite{bm93}
 to fail (or else to require recondite phase tunings
\cite{fo02}),  it was recently shown\cite{bsv03} to naturally predict
large neutrino (PMNS)  mixings in the
$\nu_{\mu}-\nu_{\tau}$ sector  due to
 $b-\tau$ unification combined with
 Type-II seesaw\cite{mohsen} domination. This natural
 solution of  the vexed question of the conflict between
  large leptonic and small quark  mixings in the
   context of GUTs (where both
  belong to common representations and
  thus often tend to have similar
   mixings)  triggered an   intensive investigation
   of the full fermion mass fitting
   problem  in such FM systems \cite{moh,bsv04,bert}
   which enjoys continuing success.

The identification of a minimal and fully realistic
model naturally motivates the development of
techniques for the  calculation of all
 mass spectra and couplings necessary for the full
  specification of the low energy effective field theory
  implied by this GUT.  The calculation of mass spectra and RG
  analysis of this GUT was begun in
  \cite{aulmoh}. It was continued in \cite{lee}
  using a somewhat abstract method\cite{heme} for calculating the
  $SO(10)-G_{123}$ clebsches relevant for mass spectra.
   However the spectra
  were quite incomplete and no method for
  evaluating the MSGUT couplings in an MSSM basis
  was available until the development of analytic
  methods to translate between SO(10) labels and
  those of its maximal subgroups $G_{PS}, SU(5)\times U(1) $
  \cite{alaps},\cite{nathraza}.
  The computation of these spectra
  was then begun by the authors of \cite{alaps} and the
  crucial ($4\times 4$ and $5\times 5$)
  mixing matrices   for the MSSM type Higgs doublets
  and proton  decay mediating triplets  were published in
 a later version \cite{alaps} of this paper. In 2004
 two independent calculations of the chiral SO(10) tensor
 mass spectra\cite{fikmo0401},\cite{bmsv04}
 and of the complete chiral and gauge
 spectra and couplings (including spinors)\cite{msgt04}
 appeared. The  results of \cite{bmsv04} and \cite{msgt04}
 had been cross checked and found
 compatible before publication. On the other hand, notable
 discrepancies existed between the results of
version 1 of  \cite{fikmo0401} and
\cite{alaps,bmsv04,msgt04}
and were pointed out in these papers. The
subsequent version of \cite{fikmo0401} then appeared,
incorporating these corrections without any
acknowledgement,  but alleging that the results of
\cite{alaps,bmsv04} were inconsistent since they
would  not satisfy the consistency tests based on Traces,
``Hermiticity" and Goldstone counting applied
 by these authors to their own results. However
  these authors did not provide any details or claim that
  they had adjusted for the phase convention differences
   between   the different calculations. There was no
   reason at all to expect the results with different
 conventions to satisfy the the consistency conditions
 in the particular form valid for the particular
 conventions they had chosen to use.
 Moreover both the impugned calculations
    had checked the Goldstone-super-Higgs
     counting down to the smallest
relevant little  group namely $G_{123}$ so there
 was no reason why the MSSM spectra should
  fail reassembly at the special vevs
  with SU(5) or Pati-Salam symmetry
   (from which decomposition of SO(10)
   the results of \cite{alaps,msgt04}
    were derived ).
     Nevertheless these authors insisted that
the other calculations
failed their tests and hence theirs were the only
correct results. In \cite{procviet} the present
author mentioned and in version 2 of \cite{msgt04}
explicitly demonstrated that all mass terms
grouped into SU(5) invariant form when the vevs were SU(5)
invariant. No assumption about zero vev of the SU(5)
 singlets in ${\bf{126,\oot}}$ were made and
 the Goldstone counting was also correct.
 Now these authors have again appeared in print
 \cite{fikmo0412} making mistaken but strong allegations
 concerning the results of \cite{alaps,bmsv04,msgt04}.
 In particular they again claim that our
 spectra fail the trace and ``Hermiticity" tests
 and although they concede that the results {\it{are}}
consistent for    $SU(5)\times U(1)_X $ symmetry
 they   claim that our results  fail Goldstone counting
   when  $U(1)_X$  is broken by non zero (SU(5) invariant) ${\bf{126,\oot}}$
 vevs.  In this communication we show explicitly
  that all these claims are false and that our
  spectra obey the Trace,``Hermiticity" and
  Goldstone counting tests when due account is taken of the
  phase conventions and the
  complete SU(5) invariant mass terms derivable from our
  results\cite{alaps,msgt04} without any change or modification whatsoever.
This should finally lay this unfortunate controversy
to rest.

\section{Summary of the MSGUT}
The  chiral supermultiplets  of the    $SO(10)$ MSGUT
 consist of the ``AM type''  totally antisymmetric tensors:
  $ {\bf{210}}(\Phi_{ijkl}),
{\bf{\overline{126}}}({\bf{\Sigb}}_{ijklm}),$
 ${\bf{126}} ({\bf\Sigma}_{ijklm})(i,j=1...10)$
which   break the GUT symmetry
 to the MSSM, together with Fermion mass (FM) Higgs
{\bf{10}}-plet(${\bf{H}}_i$).
  The  ${\bf{\overline{126}}}$ plays a dual or AM-FM
role since  it also enables the generation of realistic charged
fermion      neutrino masses and mixings (via the Type I and/or
Type II mechanisms);
 three  {\bf{16}}-plets ${\bf{\Psi}_A}(A=1,2,3)$  contain the
matter  including the three conjugate neutrinos (${\bar\nu_L^A}$).

 The   superpotential   is  the sum of $W_{GH}$
 and $W_{FM}$ :
 \begin{widetext}
 \bea
  W_{GH} &=&{1\over{2}}M_{H}H^{2}_{i} + {m \over
{ 4! }} \Phi_{ijkl}\Phi_{ijkl}+{\lambda \over
{{4!}}}\Phi_{ijkl}\Phi_{klmn} \Phi_{mnij}+{M \over { 5!
}}\Sigma_{ijklm}\overline\Sigma_{ijklm}\nonumber\\
 &+&{\eta \over
4!}\Phi_{ijkl}\Sigma_{ijmno}\overline\Sigma_{klmno}+
 {1\over{4!}}H_{i}\Phi_{jklm}(\gamma\Sigma_{ijklm}+
\overline{\gamma}\overline{\Sigma}_{ijklm}) \nonumber \\
  W_{FM}&=& h_{AB}'\psi^{T}_{A}C^{(5)}_{2}\gamma_{i}\psi_{B}H_{i}+ {1\over
5!}f_{AB}' \psi^{T}_{A}C^{(5)}_{2}{\gamma_{i_{1}}}...
{\gamma_{i_5}}\psi_{B}\overline{\Sigma}_{i_1...i_5}
 \label{W}
 \eea
\end{widetext}
 In all the MSGUT has exactly 26 non-soft parameters
 \cite{abmsv03}. The
MSSM also has 26 non-soft couplings
 so the 15 parameters of $W_{FM}$ must be
essentially responsible for the 22 parameters
 describing fermion
masses and mixings in the MSSM.

The GUT scale vevs that break the gauge symmetry down to the SM
symmetry are {\cite{aulmoh,ckn}}
${\langle(15,1,1)\rangle}_{210}:\langle{\phi_{abcd}}\rangle={a\over{2}}
\epsilon_{abcdef}\epsilon_{ef},
\langle(15,1,3)\rangle_{210}~:~\langle\phi_{ab\ta\tb}\rangle=\omega
\epsilon_{ab}\epsilon_{\ta\tb},
  \langle(1,1,1)\rangle_{210}~: ~\langle\phi_{ {\tilde
\alpha}{\tilde \beta} {\tilde \gamma}{\tilde \delta}}
\rangle=p\epsilon_{{\tilde \alpha} {\tilde \beta} {\tilde
\gamma}{\tilde \delta}},
  \langle{\overline\Sigma}_{\hat{1}\hat{3}\hat{5}
\hat{8}\hat{0}}\rangle= \bar\sigma,
\langle{\Sigma}_{\hat{2}\hat{4}\hat{6}\hat{7}\hat{9}} \rangle=
\sigma $. The vanishing of the D-terms of the SO(10) gauge sector
 potential imposes only the condition $
 |\sigma|=|{\overline{\sigma}}| $.
Except for the simpler cases corresponding to enhanced unbroken
symmetry  ($SU(5)\times U(1), SU(5), G_{3,2,2,B-L}, G_{3,2,R,B-L}$
etc)\cite{abmsv03,bmsv04} this system  of equations is essentially
cubic and can be reduced to  a single cubic   equation \cite{abmsv03}
 for a variable $x= -\lambda\omega/m$, in terms of
 which the vevs $a,\omega,p,\sigma$ are specified.

Using the above vevs and the methods of \cite{alaps} we calculated
\cite{alaps},\cite{msgt04} the complete   gauge and chiral
 multiplet GUT scale spectra
{\it{and}} couplings for the  various
MSSM multiplet sets, of which the chiral multiplets belong
 to 26 different MSSM multiplet
types of which 18 are
unmixed while the other 8 types occur in multiple copies and mix
via upto 5 x 5 matrices. For immediate reference the
chiral and chiral-gauge
 mass spectraof \cite{msgt04} are listed in  Appendix A. In the pure chiral
 sector they   are the values of
 the   mass matrices defined as the double derivatives of the
 superpotential :

 \be {\cal M}_{ij} \ =\
\frac{\partial^2 W}{\partial
 \varphi_i \partial \overline{\varphi}_j} \Bigg|_{VEV}
\ee

where $\varphi_i$ represents any $G_{321}$ multiplet  and
$\bar\varphi_1$ is its conjugate {\it{from
 the same SO(10) representation if the representation is real
 (i.e 10,210) and from the conjugate SO(10) representation
  if complex}} (i.e if $\varphi_i \in 126$  then
  $\bar\varphi_i\in \oot $ and vice versa).

\section{Consistency checks}

The basic fact about the results of the three extant
calculations of MSGUT mass spectra \cite{alaps},\cite{bmsv04,fikmo0405,
msgt04} is that all three calculations yield identical
combinations of superpotential parameters in every
single matrix element and differ only overall
phases if at all. These phase or sign differences
arise from the different field phase conventions
used and must be accounted for explcitly when
applying the consistency tests of \cite{fikmo0405}
to the other two calculations. We trace the origin and
influence of these phases and then show that there
are no inconsistencies once the effects of these phases are
accounted for properly  : as required by elementary
linear algebra.

The consistency tests insisted upon by the authors of
\cite{fikmo0412} are described below in their own words
(followed by our explanatory comments):

\subsection{Trace  consistency}
``There are three main consistency checks.
The first is that the trace of the total Higgs mass matrix does
not depend on the coupling constants $\lambda_i,\ i=1,2,3,4$. It
depends only on mass parameters $m_i,\ i=1,2,3$ and the dimensions
of the corresponding $SO(10)$ representations. The sum rule for
the Higgs-Higgsino mass matrices is "

\be ``{\rm Tr}{\cal M}\ =\
2m_1\times 210  +  m_2\times 252 +  2m_3\times 10 \;'' \ee

  Where the mass parameters  are related to ours by
  $m_1=m,m_2=2 M,m_3=M_H/2$.

 The origin of this sum rule is as follows .
 For   the {\bf{ (10)}}-plet : $H_i,1=1...10; $
  we  have

  \be Trace_{10} [{\cal M}_{ij}]   =\sum_i
\frac{\partial^2 W}{\partial  H_i \partial  H_i }
\Bigg|_{VEV} = 10 M_H
\ee

Similarly for  210 we can take the 210 independent field
 components in the real SO(10) vector label  basis to be
 $\{\Phi_{ijkl} ; i<j<k<l\}$  so that

\begin{widetext}

  \bea Trace_{210} [{\cal M}_{ij}] \ &=&
  \sum_{i<j<k<l}
\frac{\partial^2 W}{\partial \Phi_{ijkl}
 \partial \Phi_{ijkl} }\Bigg|_{VEV} \nonumber\\
 &=& 210 \times 2 m + 672\lambda
\Phi_{ijij} = 420 m
  \eea

Similarly

\bea
 Trace_{126} [{\cal M}_{ij}] \ &=&
2\sum_{{i<j<k<l<m}\in {\cal I}} \
\frac{\partial^2 W}{\partial \Sigma_{ijklm} \partial
\Sigb_{ijklm}}
\Bigg|_{VEV}\nonumber  \\&=&    504 M
\eea
here ${\cal I}$ is one of the independent
index types $abcde,\  {\bf{or}} \
 abcd\tb,\  {\bf{or}} \  abc\ta\tb \ ;
 \  a,b=1,2,3,4,5,6; \
 \ta,\tb=7,8,9,10 $ in the notation of \cite{alaps}.

So in all

\bea Trace [{\cal M}_{ij}]  &=&  10 M_H +
 420   m + 504 M   \label{tracenos}\eea

\end{widetext}

Now when taking the trace {\it{one must transform from real}}
SO(10) {\it{labels}} $i,j,k..$
 {\it{to (in general) complex}} $G_{123} $ {\it{labels}}
$\alpha_R=1,2... dim(R) $ for every $G_{123}$
sub-representation R of the relevant SO(10) tensor
representation by using a Unitary transformations  :

\bea \varphi_{\alpha_R} &=&
U_{\alpha_R, ij....}\varphi_{ij...\in R} \nonumber \\
\bar \varphi_{\alpha_{\bar R}} &=&
{\overline{U}}_{\alpha_{\bar R}, ij....}\bar\varphi_{ij...\in \bar R}
\eea
Note that overall relative phases of
 $U,\overline U$ are conventional.
Therefore the trace in $G_{123}$ labels is really
to be done with the metric

\be (U{\overline{U}}^T)_{\alpha_R,  \beta_{\bar R}} \ee

{\it{This metric is not necessarily unit}}.
  In fact in our basis\cite{alaps} which is derived from an
embedding of $G_{PS}$ in SO(10)\cite{alaps}
its diagonal elements  can be $\pm 1 $  and
 this results precisely in negative signs for some of the
 mass parameter terms in the diagonal mass matrix elements.
  Its value is thus obviously the sign of
the mass parameter in the diagonal mass terms for each
$G_{123}$ subrep, so that the minus signs that occur
(e.g for ${\bar C}_2 C_2,{\bar E_1} E_1 $
etc :  see the Appendix) are automatically
 compensated to ensure that at least
 the mass parameter terms
 give the correct contributions to
 eqn.(\ref{tracenos}). The consistency check is
 then to see whether {\it{after compensating
 for these minus signs }} the coupling
 parameter ($\lambda,\eta$) dependent terms
 indeed add up to zero (since they are proportional to
 $ \Phi_{ijij}$ )in both cases.
  It is gratifying to check  that
the sum of   some hundred different terms
 added up to precisely zero thus verifying
 eqn.(\ref{tracenos}). We have given the details of this
 trace calculation in the Appendix in the form of
  two tables. It thus appears that the authors of
 \cite{fikmo0412} did not account for these elementary
 facts of linear algebra before claiming that our phases
 were inconsistent.
 In short the Trace constraint is in fact satisfied
 by our spectra just as it is by theirs.

    \subsection{SU(5) Check}

    Next  we come to what these authors consider their
    ``main check ''  but to us  seems a trivial
    consequence
    of the little group embedding
    $G_{123}\subset SU(5) $ once
    one has checked the super-Higgs effect for the
    breaking $SO(10) \rightarrow G_{123}$ (as we have
    done in in great and explicit detail including
    calculation of the null eigenvectors)\cite{msgt04}.
    However these authors make the alarming and bald
    assertion :

   ``The third and main check is the $SU(5)$ check briefly
   described in the paper \cite{fikmo0405},
 but misunderstood in Refs. \cite{msgt04,procviet}.
  Here we explicitly prove
  that mass matrices in Ref. \cite{fikmo0401} satisfy
  this highly nontrivial test, and that  the
   results obtained in Refs. \cite{bmsv04,alaps,msgt04}
      are internally inconsistent. "
      (our reference numbers).

 We had already\cite{procviet},\cite{msgt04}(version 2)
   shown that this SU(5)
 re-assembly is possible .
  Indeed the relevant mass  terms( for $p=a=-\omega$)
   were given explicitly
   there :
 \begin{widetext}
 \begin{eqnarray}
  && 2(M  +  10\eta p) 1_{\Sigma} 1_{\overline\Sigma} + 2(M + 4 \eta p)
  \bar 5_{\Sigma}  5_{\overline\Sigma}  + 2(M - 2\eta
p) \overline{50}_{\Sigma} 50_{\overline\Sigma} \nonumber \\
&+& 2(M + 4\eta p) 10_{\Sigma}
\overline{10}_{\overline\Sigma} + 2(M + 2\eta p)
\overline{15}_{\Sigma}  15_{\overline\Sigma}
+ 2M 45_{\Sigma} \overline{45}_{\overline\Sigma}  \nonumber \\
 & + & M_H  \bar 5_H  5_H + ( m + 6\lambda p) (1_{\Phi})^2 +
 2(m + 6 \lambda p) 5_{\Phi} \bar 5_{\Phi} + 2(m
  + 3  \lambda p) 10_{\Phi} \overline{10}_{\Phi} \nonumber \\
 &+& (m + \lambda p)
 (24_{\Phi})^2 + 2m 40_{\Phi} \overline{40}_{\Phi} + (m - 2\lambda p)
 (75_{\Phi})^2 \nonumber \\
  &+& 2\eta\sqrt 3 (\overline\sigma (\bar 5_{\Sigma}
 5_{\Phi} + 10_{\Sigma} \overline{10}_{\Phi})
  +  \sigma ( 5_{\overline\Sigma} \bar 5_{\Phi}
  + \overline{10}_{\overline\Sigma} 10_{\Phi}))  \nonumber \\
 &+& 2\sqrt 3 p(\gamma \bar 5_{\Sigma} 5_H +
 \bar{\gamma} \bar 5_H 5_{\overline \Sigma})
+ 2i {\eta\sqrt 5} ( \sigma  1_{\overline\Sigma} 1_{\Phi}
 - \overline\sigma  1_{\Sigma}1_{\Phi})\label{su5reass}
\end{eqnarray}
  \end{widetext}
Note that the  trace requirement eqn.(\ref{tracenos}) is satisfied.
 The last term involving only singlets  is
absent when $\sigma=\bar\sigma =0$ and was not listed
when detailing the $SU(5) \times U(1)$ reassembly
 \cite{msgt04},  but it  is of course
always present in the  $G(1,1,0)$ sector
 mass matrix since the mixing terms between
  $G_4= 1_{\Sigma},  G_5= 1_{\Sigb}$ and $G_1,G_2,G_3$
  are read off from ${\cal G}_{(1,1,0)}$(see the Appendix)
   to be

 \bea &&  i {\eta\sqrt 2} (\ssb G_4 - \sigma G_5 )( G_1 +
 \sqt G_2 -\sqs G_3)\nonumber \\
 &=& 2i {\eta{\sqrt 5}} ( \sigma  1_{\overline\Sigma} 1_{\Phi}
 - \overline\sigma  1_{\Sigma}1_{\Phi})\nonumber\\
 1_{\Phi} &=& -\sqrt{1\over 10} ( G_1 +
 \sqt G_2 -\sqs G_3)
 \eea

From these mass terms one can easily verify that
the SU(5) singlet mass matrix with rows
labelled by $(1_{\Sigma},1_{\Sigb}, 1_{\Phi})$
and columns by $(1_{\Sigb},1_{\Sigma}, 1_{\Phi})$

is just

\bea {\cal{M}}_1= \left({\begin{array}{ccc}
  2(M+10\eta p)&0&-2i {\eta\sqrt 5}\ssb\\
 0&2(M+10\eta p) &2i {\eta\sqrt 5} \sigma\\
-2i {\eta\sqrt 5}\ssb &2i {\eta\sqrt 5}\sigma&2(m + 6 \lambda p)
 \end{array}}\right)
\eea

whose determinant is manifestly zero at
the SU(5) symmetric point  where $p=-M/10\eta$ but is
{\it{not}} zero at the generic $SU(5)\times U(1) $
invariant vevs. Thus we obtain the SU(5) singlet Goldstone
required in the former case.

Moreover the $10-{\overline {10}}$ sector matrix is
trivially read off from the terms given above to be

\bea {\cal{M}}_{10} = \left({\begin{array}{ccc}
   2(M+4\eta p)&2 {\eta\sqrt 3}\sigma\\
{2\eta\sqrt 3} \ssb &2 (m+ 3\lambda p)
  \end{array}}\right)
\eea

The determinant at the SU(5) solution point
 $ p=-M/{10\eta},\quad  \sigma\ssb=-2p(m+3\lambda p )/{\eta} $
 is again zero. This shows that in fact, just as found by
 the authors of \cite{fikmo0405} there are at most
  13 distinct mass eigenvalues namely

  \bea (m_1)_1= 0 \quad; \qquad
  (m_{(10 +{\overline{10}})})_1=0  \nonumber \\
  {\bf{2}} (m_1)_{2,3} ; \qquad
  (m_{(10 +{\overline{10}})})_2 \nonumber\\
   {\bf{3}} (m_{(5+\bar 5)})_{1,2,3} ; \qquad
   m_{(50 +{\overline{50}})} \nonumber\\
     m_{(40 +{\overline{40}})} ;\qquad
         m_{(15 +{\overline{15}})} \nonumber\\
    m_{(45 +{\overline{45}})} ;
    \qquad (m_{24});\quad (m_{75})
 \eea

The trace of this mass matrix is also easily found to
be the same as eqn(\ref{tracenos}). It is is also
 easy to check that after accounting for the
 trace metric, as explained above, and up
  to a phase in the case of the determinant
   eqn.(17) of \cite{fikmo0412} namely :

 \bea
 \label{DTTrDet}
 {\rm tr} \; {\cal T}_{\sf triplet} &=&
 m_\Sigma({\bf 50}) + {\rm tr}\; {\cal H}_{\sf doublet}\;,
 \nonumber\\
 {\rm det}\; {\cal T}_{\sf triplet} & =&
 m_\Sigma({\bf 50}) \times {\rm det}\; {\cal H}_{\sf doublet}\;.
 \eea
are obeyed at the SU(5) point and
 $ m_\Sigma({\bf 50})=2 (M-2 \eta p) = 12 M/5 $.
 Note   that in this case the
trace and determinant has been taken without weighting for the
dimension of the doublet and triplet
representations.

 To sum up , the authors of \cite{fikmo0412} had already
 conceded that our results  were consistent when
 $\sigma=\bar\sigma=0$ but asserted that we had
  misunderstood their consistency tests for the case with
  only SU(5) symmetry where
    $ p=-M/{10\eta},\quad
    \sigma\ssb=-2p(m+3 \lambda p )/{\eta} $
     and that our results failed the counting of Goldstone
      multiplets which restricted the number of independent
      mass eigenvalues  to just 13.
Yet we have just shown, without any new result whatever that
    the SU(5) solution is just as consistent as the
    $SU(5)\times U(1)$ solution. As we noted earlier, in view
     of the fact that we have already analyzed
     the mass matrices  -and found them consistent-
 for  the breaking down to the much smaller little group
$G_{123}$, the reassembly into $SU(5) , SU(5)\times U(1)$ or
$G_{PS}$ invariants when the vevs have these symmetries
  is no surprise at all. The wonder of these
  authors at this phenomenon is perhaps
  attributable to the extreme formality
  of their calculational approach in which simple
  algebraic facts take on a
   mathematical sheen difficult to demystify.

\subsection{``Hermiticty" Property }

The next check proposed by these authors is the so-called
``Hermiticity" cross check. Properly speaking, since we
are calculating chiral mass matrices ``hermiticity"
of the mass matrices is an oxymoron. However
by adopting particular phase conventions and
considering the sum of invariants involving
 $ 126$  and $\oot $ separately  to be ``real''
these authors obtain a relation between their Clebsches
that they call a ``hermiticity'' relation though no
actual complex conjugation is involved.

The corresponding symmetry from our point of
view is the observation that our GUT Higgs
superpotential   $W_{GH} $ (not $W_{FM}$ )
is invariant under the substitutions

\be \Sigma \longleftrightarrow  \Sigb \quad ;
\quad \gamma    \longrightarrow  \bar\gamma
\ee

and the $10,210$ representations are real.

Thus if we wrote all our mass matrices
 in terms of the original real labels and  kept
  track of any sign or phase introduced while defining vevs
 of  SO(10) tensors then
 we would indeed observe the effects of this symmetry as
 a transposition property of the mass matrices. In fact
  it is apparent that our mass matrices obey this
  transposition symmetry {\it{up to  signs}}. However
 after transforming\cite{alaps}
  to our PS derived complex labels each off diagonal
  mass term contains {\it{three}} phases
  introduced by this transformation. The relative signs
  which need to be checked are thus a ratio of {\it{six}}
   phases !   To trace back and recalculate
   all the phases involved
  in some hundreds of  terms merely to
  satisfy the contrivance of an independent
  phase convention would negate the entire spirit of our
  method which is  based on keeping automatic  track of
  these phases by taking the trouble to perform
   the embedding of PS in SO(10) for all the relevant
   representations completely explicitly.
   Thus if these authors wish to insist that
   our results fail this contrived test
   formulated in terms of Clebsch coefficients
    that we never define or use  it is incumbent on them
    to trace these 3 phases for every off
    diagonal mass term back to the real
    labels and show that the results conflict with this
hidden  ``symmetry". This is particularly so
 after our demonstration that all their other assertions are
 false due to elementary confusions.
   Since all other computable tests
are in fact obeyed an appeal to this obscure requirement
would seem to clutch at straws !

\section{Conclusions }
The detailed  discussion
above demonstrates that our mass matrices
(and by implication those of
\cite{bmsv04})   satisfy   every reasonable
consistency check claimed by the authors of
\cite{fikmo0405,fikmo0412} to be {\it{unique}}
to their calculations and  makes it clear that these three
calculations are equivalent as far as Chiral mass
spectra are concerned. However the method of \cite{alaps}
yields not only chiral mass spectra and chiral vev equation
decompositions in terms of MSSM labels but also all other
couplings whether gauge or chiral and for both tensor
and spinor SO(10)representations. The methods of
\cite{heme} used in the calculations of
\cite{bmsv04,fikmo0405} cannot give
 these quantities even in principle.   Our
method is different from the  formal method
of\cite{heme,lee, bmsv04,fikmo0405} and
is more complete, especially regarding
couplings. Being analytic and explicit it
also allows us to trace and resolve
discrepancies in a form directly
usable in a field theory calculation .
 Since the excavation of the phenomenological
   implications of
 the MSGUT requires knowledge of most of these quantities
 it seems that use of the phase conventions of
 \cite{fikmo0405}  will carry the burden of an additional
 and very tedious phase accounting before our results
 on the most general
 couplings\cite{msgt04} -the only ones available-
  are usable under alternate conventions.

 \vspace{ .5 true cm}
 {\bf {Appendix   : Tables
of masses and mixings }} \vskip .5 true cm
 \vspace{ .2 true cm}

In this appendix we give the mass spectra \cite{msgt04}
 of the chiral fermion/gaugino states together with
  the Trace calculation in MSSM labels.
   Mixing matrix rows are labelled by barred
  irreps and columns by unbarred.
  Unmixed cases({\bf{i)}}) are given as Table
  I.
\begin{widetext}
\begin{table}
$$
\begin{array}{l|l|l|l}
{\rm Field }[SU(3),SU(2),Y] &  PS
\qquad  Fields  & {\rm Multiplicity \times Mass}& Metric  \\
 \hline
 &&\\
 A[1,1,4],{} \bar A[1,1,-4] &{{\s^{44}_{(R+)}}\over
\sq}, {{\os_{44(R-)}}\over \sq}&
 4( M + \eta (p +3a +  6 \omega ))&+ \\
 C_1[8,2,1],{} \bar C_1 [(8,2,- 1] &
\s_{\bn\alpha\dot1}^{~\bar\lambda},{}
 \Sigb_{\bn\alpha\dot 2}^{~\bar\lambda} &
 64(-M + \eta (a+\omega))&- \\
 &&\\
C_2[8,2,1],{}\bar C_2 [(8,2,- 1] &
\Sigb_{\bn\alpha\dot1}^{~\bar\lambda},{}
 \s_{\bn\alpha\dot 2}^{~\bar\lambda} &
 64 (-M + \eta (a-\omega))&- \\
D_1[3,2,{7\over 3}],{}\bar D_1 [(\bar 3,2,- {7\over 3}] &
\Sigb_{\bn\alpha\dot1}^{~4},{}
 \s_{4\alpha\dot 2}^{~\bn} &
  24 (M + \eta (a+\omega)) &+\\
 &&\\
D_2[3,2,{7\over 3}],{}\bar D_2 [(\bar 3,2,- {7\over 3}]
 & \s_{\bn\alpha\dot1}^{~4},{}
 \Sigb_{4\alpha\dot 2}^{~\bn} &
  24 (M + \eta (a+3\omega)) &+\\
 &&\\
E_1[3,2,{1\over 3}],{}\bar E_1 [(\bar 3,2,- {1\over 3}] &
\Sigb_{\bn\alpha\dot 2}^{~4},{}
 \s_{4\alpha\dot 1}^{~\bn} &
 -  24 (M + \eta (a - \omega)) &-\\
K[3,1,-{8\over 3}],{}\bar K [(\bar 3, 1, {8\over 3}] & \Sigb_{\bn
4(R-)},{}
 \s^{\bn4}_{(R+)} &
 12 (M + \eta (a+p+ 2 \omega)) &+\\
 &&\\
L[6,1,{2\over 3}],{}\bar L [(\bar 6,1, -{2\over 3}] &
  (\Sigb_{\bm\bn}^{'(R0)},
 \s^{'\bm\bn}_{(R0)})_{\bm\leq\bn} &
 24 (M + \eta (p -a)) &+\\
& \Sigb'_{\bm\bn}=\Sigb_{\bm\bn},{} \bm\neq \bn&\\
&\Sigb'_{\bm\bm}={{\Sigb_{\bm\bm}}\over\sq}&\\
 &&\\
   M[6,1,{8\over 3}],{}{\ovl M} [(\bar 6,1, -{8\over 3}] &
  (\Sigb^{'(R+)}_{\bm\bn(R+)},{}
 \s^{'\bm\bn}_{(R-)})_{\bm\leq\bn} &
 24 (M + \eta (p -a + 2 \omega )) &+\\
N[6,1,-{4\over 3}],{}\bar N [(\bar 6,1, {4\over 3}] &
 (\Sigb_{\bm\bn}^{'(R-)},
  \s^{'\bm\bn}_{(R+)} )_{\bm\leq\bn} &
 24 (M + \eta (p -a-2\omega ))&+ \\
 O[1,3,-2],{}\bar O [(1,3, +2] &
 {{{\vec\s}_{44(L)}}\over \sq},{}
 {{{{\vec{\Sigb}}^{44}_{(L)}}}\over \sq} &
 12 (M + \eta (3a-p))&+ \\
 &&\\
 P[3,3,-{2\over 3}],{}\bar P [\bar 3,3, {2\over 3}] &
 {\vec\s}_{\bm 4(L)},
  {\vec\Sigb}^{\bm 4}_{(L)}  &
 36 (M + \eta (a-p)) &+\\
  &&\\
W[6,3,{2\over 3}],{}{\overline W} [({\bar 6},3, -{2\over 3}] &
 {{{\vec\s'}_{\bm\bn(L)}}} ,
 {\vec\Sigb}^{\bm \bn}_{(L)}  &
 72 (M - \eta (a+p)) \\
I[3,1,{10\over 3}],{}\bar I [(\bar 3,1,- {10\over 3}] &
\phi_{~\bn(R+)}^4,{}
 \phi_{4(R-)}^{~\bn} &
 - 6(m + \lambda (p+a+4\omega))&- \\
S[1,3,0] & \vec\phi^{(15)}_{(L)} & 6(m+\lambda(2a-p))&+\\
Q[8,3,0]& {\vec\phi}_{\bm(L)}^{~\bn}&
  48 (m - \lambda (a +p)) &+\\
U[3,3,{4\over 3}],{} \bar U[ \bar 3,3,-{4\over 3}] &
{\vec\phi}_{\bm(L)}^{~4},{} {\vec\phi}_{4(L)}^{~\bm}&
 -36 (m - \lambda (p-a)) &-\\
 &&\\
V[1,2,-3],{} \bar V[ 1,2,3] & {{{\phi}_{44\alpha\dot
2}}\over\sq},{} {{\phi^{44}_{\alpha\dot 1}}\over \sq}&
  8 (m  + 3 \lambda (a + \omega))&+ \\
B[6,2,{5\over 3}],{}\bar B [(\bar 6,2, -{5\over 3}] &
 (\phi_{\bm\bn\alpha\dot 1}',
 \phi^{'\bm\bn}_{\alpha\dot 2} )_{\bm\leq\bn} &
 -48 (m + \lambda (\omega -a ))&- \\
Y[6,2,-{1\over 3}],{}\bar Y [(\bar 6,2, {1\over 3}] &
 (\phi_{\bm\bn\alpha\dot 2}',
\phi^{'\bm\bn}_{\alpha\dot 1})_{\bm\leq\bn} &
 48 (m - \lambda (a+\omega ))&+ \\Z[8,1,2],{} \bar Z[ 8,1,-2] & {\phi}_{~\bm(R+)}^{\bn}
{\phi}_{\bm(R-)}^{~\bn}&
 32 (m + \lambda (p-a)) &+ \\
 Sub-total&&408 M+ 238 m + &\\
 &&+\eta (-32 p -104 a + 120 \omega) &\\
 &&+\lambda (- 46 p - 92a + 72 \omega ) &
\end{array}
$$
\label{table I} \caption{{\bf{i)}}
 Multiplicity weighted Masses and trace metric signs
 for the unmixed states.   }
\end{table}

\begin{table}
$$
\begin{array}{l|l|l}
{\rm Field }[SU(3),SU(2),Y] &
 {\rm Multiplicity \times  Mass}& Metric  \\
 \hline
 E_2[3,2,{1\over 3}],{}\bar E_2 [(\bar 3,2,- {1\over 3}]
  &  -24 (M + \eta (a - 3\omega)) &-\\
E_3[3,2,{1\over 3}],{}\bar E_3 [(\bar 3,2,- {1\over 3}]
  &  -24 (m + \lambda (a - \omega)) &-\\
E_4[3,2,{1\over 3}],{}\bar E_4 [(\bar 3,2,- {1\over 3}]
  & - 24 (m -\lambda  \omega)) &-\\
 F_1[1,1,2],\bar F_1[1,1,-2]&4(M+\eta(p+ 3 a))&+\\
F_2[1,1,2],\bar F_2[1,1,-2]&4(m+\lambda(p+ 2 a))&+\\
G_1[1,1,0] &2m &+\\
G_2[1,1,0] &2(m + 2 \lambda a)&+\\
G_3[1,1,0] &2 (m + \lambda(p+2 a)) &+\\
G_4[1,1,0] &2(M+\eta (p+3 a -6 \omega)) &+\\
G_5[1,1,0] &2(M+\eta (p+3 a -6 \omega)) &+\\
h_1 [1,2,1], {\bar h}_1 [1,2,-1]&-4 M_H &-\\
h_2 [1,2,1], {\bar h}_3 [1,2,-1]&- 8(M+2 \eta(a -\omega))
&-\\
h_3 [1,2,1], {\bar h}_2 [1,2,-1]&- 8(M+2 \eta(a +\omega))
&-\\
h_4[1,2,1], {\bar h}_4 [1,2,-1]
& - 8(m + 3 \lambda (a-\omega)) &-\\
 J_1[3,1,{4\over 3}],\bar J_1[\bar 3,1,-{4\over 3}]&
 12(M+\eta(a +p-2 \omega)&+\\
J_2[3,1,{4\over 3}],\bar J_2[\bar 3,1,-{4\over 3}]&
 -12( m+\lambda  a )  &-\\
J_3[3,1,{4\over 3}],\bar J_3[\bar 3,1,-{4\over 3}]&
 -12( m+\lambda(a +p)&-\\
R_1[8,1,0]& 16(m-\lambda a) &+\\
R_1[8,1,0]& 16(m+\lambda (p-a)) &+\\
t_1[3,1,-{2\over 3}], \bt_1[\bar 3,1,{2\over 3}]&
6M_H&+\\
t_2[3,1,-{2\over 3}], \bt_3[\bar 3,1,{2\over 3}]&
12 M&+\\
t_3[3,1,-{2\over 3}], \bt_2[\bar 3,1,{2\over 3}]&
12 M&+\\
t_4[3,1,-{2\over 3}], \bt_4[\bar 3,1,{2\over 3}]&
12 (M + \eta (p+ a))&+\\
t_5[3,1,-{2\over 3}], \bt_5[\bar 3,1,{2\over 3}]&
-12 (m + \lambda (p+ a-4 \omega))&-\\
 \bar X_1[3,2,{5\over 3}]
 X_1[3,2,-{5\over 3}] \bar X_1[3,2,{5\over 3}]&
 24(m +  \lambda (a+\omega))&+\\
 X_2[3,2,-{5\over 3}] \bar X_2[3,2,{5\over 3}]&
 24(m +  \lambda  \omega )&+\\
 Sub-total& 96 M+ 182 m 10 M_H  &\\
 &+\eta ( 32 p +104 a  - 120 \omega) &\\
 &+\lambda (46 p +92 a  - 72 \omega ) &
\end{array}
$$
\label{table II} \caption{{\bf{i)}}
 Multiplicity weighted Masses and trace metric signs
 for the  mixed states. The sum of the
 Subtotals of Table I and II gives precisely
 $10 M_H + 420 m + 504 M $, with all
 coupling dependent terms cancelling  as required.   }
\end{table}

 {\bf{ ii)\hspace{ 1.0 cm} Chiral Mixed states}}\hfil\break

\vspace{ .3 cm}

 {\bf{a)}}$ [8,1,0](R_1,R_2)\equiv (\hat\phi_{\bm}^{~\bn},\hat\phi_{\bm
(R0)}^{~\bn})  $

 \be
{\cal{R}} = 2 \left({\begin{array}{cc} (m-\lambda a ) &
-\sqrt{2}\lambda\omega \\ -\sqrt{2}\lambda\omega & m+\lambda( p-a)
\end{array}}\right)
\ee

{\bf{b)}}\hspace{ 1.0cm}  $ [1,2,-1]({\bar h}_1,\bh_2,\bh_3,\bh_4)
\oplus     [1,2,1](h_1,h_2,h_3,h_4) $\hfil\break
 $.\qquad\qquad\equiv
(H^{\alpha}_{\dot 2},\Sigb^{(15)\alpha}_{\dot 2},
\s^{(15)\alpha}_{\dot2},{{\phi_{44}^{\dot 2\alpha}} \over
\sq})\oplus (H_{\alpha {\dot 1}},\Sigb^{(15)}_{\alpha \dot1},
\s^{(15)}_{\alpha\dot 1}, {{\phi^{44\dot 1}_{\alpha}} \over \sq})
 $

\bea {\cal{H}}=\left({\begin{array}{cccc} -M_{H} &
+\overline{\gamma}\sqrt{3}(\omega-a) & -{\gamma}\sqrt{3}(\omega+a)
& -{\bar{\gamma}{\bar{\sigma}} }\\
 -\overline{\gamma}\sqrt{3}(\omega+a) & 0 & -(2M+4\eta(a+\omega)) &
0\\
\gamma\sqrt{3}(\omega-a) & -(2M+4\eta(a-\omega)) & 0 &
-{2\eta\overline{\sigma}\sqrt{3}}\\
  -{\sigma\gamma } &
-{2\eta\sigma\sqrt{3}} & 0 & {-{2m}+6\lambda(\omega-a)}
\end{array}}\right) \nonumber \eea

\vspace{ .5 true cm}

 {\bf{c)}} $[\bar 3,1,{2\over 3}]
(\bt_1,\bt_2,\bt_3,\bt_4,\bt_5) \oplus [3,1,-{2\over 3}]
(t_1,t_2,t_3,t_4,t_5)$\hfil\break $.\qquad\qquad\equiv (H^{\bm
4},\Sigb_{(a)}^{\bm 4}, \s^{\bm 4
}_{(a)},\s^{\bm4}_{R0},\phi_{4(R+)}^{~\bm}) \oplus (H_{\bm
4},\Sigb_{(a)\bm 4}, \s_{\bm 4 (a)},\Sigb_{\bm
4(R0)},\phi_{\bm(R-)}^{~4}) $

\bea {\cal{T}}= \left({\begin{array}{ccccc} M_{H} &
\overline{\gamma}(a+p) & {\gamma}(p-a) & {2\sqrt{2}i
\omega{\bar\gamma}} & i\bar{\sigma}\bar{\gamma}\\ \bar\gamma(p-a)
& 0 & 2M & 0 & 0\\ \gamma(p+a) & 2M & 0 & 4\sqrt{2}i \omega\eta &
2i{\eta\overline{\sigma}}\\ -2\sqrt{2}i \omega\gamma  &
-4\sqrt{2}i \omega\eta & 0 & 2M+2\eta{p}+2\eta a &
-2\sqrt{2}{\eta\overline{\sigma}}\\ i\sigma\gamma & 2i \eta\sigma
& 0 & 2\sqrt{2}\eta\sigma & -2m - 2\lambda(a+p-4 \omega)
\end{array}}\right)
\nonumber\eea

\vspace{ .5 cm}

 {\bf{iii)  Mixed gauge chiral .}}

{\bf{a)}}$ [1,1,0] (G_1,G_2,G_3,G_4,G_5,G_6) \equiv
(\phi,\phi^{(15)},\phi^{(15)}_{(R0)},{{\s^{44}_{(R-)}}\over \sq},
{{\Sigb_{44((R+)}}\over \sq}, {{{\sq \lambda^{(R0)} -
{\sqrt{3}}\lambda^{(15)}}\over {\sqrt{5}}}})$

\bea {\cal{G}}= 2\left({\begin{array}{cccccc} m&0 &
\sqs\la\om & {{i\e\ssb}\over \sq}&{-i\e\sss\over \sq}&0\\
0& m + 2 \la a & 2\sq\la\om& i\e\ssb\sqtt &-i\e\sss\sqtt&0\\
\sqs\la\om&2\sq\la\om&m+\la(p+2a)& -i\e\sqt\ssb & i\sqt\e\sss&0\\
{{i\e\ssb}\over\sq}& i\e\ssb\sqtt&-i\e\sqt\ssb&0&
M+\e(p+3a -6\om)&{{\sqf g\sss^*}\over  2 }\\
{{-i\e\sss}\over\sq}& -i\e\sss\sqtt&i\e\sqt\sss&
M+\e(p+3a -6\om)&0&{{\sqf g \ssb^*}\over 2}\\
0&0&0&{{\sqf g\sss^*}\over 2}&{{\sqf g\ssb^*}\over 2}&0
\end{array}}\right)
\nonumber\eea

{\bf{b)}}  $[\bar 3,2,-{1\over 3}](\bar E_2,\bar E_3,\bar E_4,\bar
E_5) \oplus [3,2,{1\over 3}](E_2,E_3,E_4,E_5)$\hfil\break
$.\qquad\qquad \equiv (\Sigb_{4\alpha\dot 1}^{\bm}, \phi^{\bm
4}_{(s)\alpha\dot 2} , \phi^{(a) \bm 4}_{\alpha\dot
2},\lambda^{\bm 4}_{\alpha\dot 2}) \oplus  (\s_{\bm\alpha\dot
2}^4,\phi_{\bm 4\alpha\dot 1}^{(s)}, \phi_{\bm 4\alpha\dot
1}^{(a)},\lambda_{\bm\alpha\dot 1}) $

\bea {\cal{E}}= \left({\begin{array}{cccc} -2(M+\e(a-3\om))&
-2\sq i\e\sss&2i\e\sss&ig\sq\ssb^*\\
2i\sq\e\ssb&-2(m+\la(a-\om))&-2\sq\la\om&2g(a^*-\om^*)\\
-2i\e\ssb&-2\sq\la\om&-2(m-\la\om)&\sq g(\om^*-p^*)\\
-ig\sq\sss^*&2g(a^*-\om^*)&g\sq(\om^*-p^*)&0
 \end{array}}\right)
\eea

{\bf{c)}}$ [1,1,-2](\bar F_1,\bar F_2, \bar F_3)
 \oplus [1,1,2](F_1,F_2,F_3) $ \hfil\break
$.\qquad\qquad \equiv
(\Sigb_{44(R0)},\phi^{(15)}_{(R-)},\lambda_{(R-)})
  \oplus (\s^{44}_{(R0)},\phi^{(15)}_{(R+)},\lambda_{(R+)}) $ .

\bea {\cal{F}}= \left({\begin{array}{ccc} 2(M+\e(p+3 a))&
-2i\sqt\e\sss&-g\sq\ssb^*\\
2i\sqt\e\ssb&2(m+\la(p+2a))& {\sqrt{24}}ig\om^*)\\
 -g\sq\sss^*&-{\sqrt{24}}ig\om^*&0
 \end{array}}\right)
\eea

{\bf{d)}} $[\bar 3,1,-{4\over 3}](\bar J_1,\bar J_2,\bar J_3,\bar
J_4) \oplus [3,1,{4\over 3}](J_1,J_2,J_3,J_4)$ \hfil\break
$.\qquad\qquad\equiv (\s^{\bm4}_{(R-)},\phi_4^{\bm},
\phi_4^{~\bm(R0)},\lambda_4^{~\bm}) \oplus
(\Sigb_{\bm4(R+)},\phi_{~\bm}^4,
\phi_{\bm(R0)}^{~4},\lambda_{\bm}^4)$

\bea {\cal{J}}= \left({\begin{array}{cccc} 2(M+\e(a+p-2\om))&
-2\e\ssb&2\sq \e\ssb&-ig\sq\sss^*\\
2\e\sss&-2(m+\la a)&-2\sq\la\om&-2ig\sq a^*\\
-2\sq\e\sss&-2\sq\la\om&-2(m+\la(a+p))&-4i g\om^*\\
-ig\sq\ssb^*&2\sq ig a^*&4i g\om^*&0
 \end{array}}\right)
\eea

{\bf{e)}}$ [3,2,{5\over 3}](\bar X_1,\bar X_2,\bar X_3) \oplus
[3,2,-{5\over 3}](X_1,X_2,X_3)\hfil\break .\qquad\qquad \equiv
(\phi^{(s)\bm4}_{\alpha\dot 1} , \phi^{(a)\bm4}_{\alpha\dot 1}
,\lambda^{\bm4}_{\alpha\dot 1}) \oplus(\phi_{\bm4\alpha\dot
2}^{(s)}, \phi_{\bm4\alpha\dot 2}^{(a)},\lambda_{\bm4\alpha\dot
2})
  $

\bea {\cal{X}}= \left({\begin{array}{ccc} 2(m+\la(a+\om))&
-2\sq \la \om &-2g(a^*+\om^*)\\
-2\sq \la \om &2(m+\la \om)& {\sq}g(\om^* +p^*)\\
 -2 g(a^* +\om^*) &\sq g(\om^* + p^*)&0
 \end{array}}\right)
\eea

\end{widetext}


\begin{thebibliography}{99}
\bibitem{msgt04} C.S. Aulakh and A. Girdhar,
 arXiv: hep-ph/0405074, to appear in Nuc. Phys. B.
\bibitem{aulmoh}
C.S. Aulakh and R.N. Mohapatra,
 Phys. Rev. {\bf D28}, 217 (1983);
\bibitem{ckn}T.E. Clark, T.K. Kuo, and N. Nakagawa,
 Phys. Lett. {\bf B 115}, 26 (1982);
\bibitem{abmsv03} C.S. Aulakh, B. Bajc, A. Melfo, G. Senjanovi\'c and F. Vissani,
Phys. Lett. {B 588}, 196 (2004).
\bibitem{lee} D.G. Lee, Phys. Rev. {\bf{D49}}, 1417 (1995).
 \bibitem{heme} X.G. He and S. Meljanac,
 Phys. Rev. {\bf{D41}}, 1620 (1990).
\bibitem{bm93}
K.S. Babu and R.N. Mohapatra,
 Phys. Rev. Lett. {\bf 70}, 2845 (1993).
  \bibitem{fo02}
T. Fukuyama and N. Okada, JHEP {\bf 0211}, 011 (2002); K. Matsuda,
Y. Koide, T. Fukuyama and H. Nishiura, Phys. Rev. {\bf D65},
033008 (2002) [Erratum-ibid. {\bf D65}, 079904 (2002)]; K.
Matsuda, Y. Koide and T. Fukuyama, Phys. Rev. {\bf D64}, 053015
(2001).
\bibitem{mohsen}
R.N.~Mohapatra and G.~Senjanovi\'c, Phys. Rev. {\bf D23},165
(1981); G. Lazarides, Q. Shafi and C. Wetterich, Nucl. Phys.
{\bf{B181}}, 287 (1981).
\bibitem{bsv03}
B. Bajc, G. Senjanovi\'c and F. Vissani, Phys.\ Rev.\ Lett.\  {\bf
90}, 051802 (2003) [arXiv:hep-ph/0210207]
\bibitem{moh} H.S.Goh, R.N.Mohapatra and S.P.Ng, Phys. Lett. B {\bf 570} (2003)215.
H.S.Goh, R.N.Mohapatra and S.P.Ng, Phys. Rev. D {\bf
68}(2003)115008.
\bibitem{bsv04}
B.~Bajc, G.~Senjanovic and F.~Vissani,
Phys.\ Rev.\ D {\bf 70}, 093002 (2004)
[arXiv:hep-ph/0402140].
\bibitem{bert}
S.~Bertolini, M.~Frigerio and M.~Malinsky,
Phys.\ Rev.\ D {\bf 70}, 095002 (2004)
[arXiv:hep-ph/0406117].
\bibitem{alaps} C.S.Aulakh and A. Girdhar, {\it{SO(10) a la
Pati-Salam }},  hep-ph/0204097; v2 August 2003;
 v4, 9 February, 2004; Intn'l J. Mod. Phys. A  (in press).
\bibitem{nathraza}
P.~Nath and R.~M.~Syed,
Phys.\ Lett.\ B {\bf 506}, 68 (2001)
[Erratum-ibid.\ B {\bf 508}, 216 (2001)]
[arXiv:hep-ph/0103165].;
P.~Nath and R.~M.~Syed,
Nucl.\ Phys.\ B {\bf 618}, 138 (2001)
[arXiv:hep-th/0109116].
\bibitem{fikmo0401}
T. Fukuyama, A. Ilakovac, T. Kikuchi, S. Meljanac, and N. Okada,
arXiv: hep-ph/0401213.
\bibitem{bmsv04}
B. Bajc, A. Melfo, G. Senjanovi\'c and F. Vissani, Phys. Rev. {\bf
D 70}, 035007 (2004).
\bibitem{fikmo0405} Fukuyama, A. Ilakovac, T. Kikuchi,
 S. Meljanac, and N. Okada, arXiv: hep-ph/0405300.
 \bibitem{fikmo0412} T. Fukuyama, A. Ilakovac, T. Kikuchi, S. Meljanac, and N. Okada,
 {hep-ph/0412348}.
\bibitem{procviet}
 C.~S.~Aulakh,
arXiv:hep-ph/0410308.



 \end{thebibliography}
 \end{document}